# UPGRADES OF THE SPS, TRANSFER LINE AND LHC INJECTION PROTECTION DEVICES FOR THE HL-LHC ERA

Ö. Mete[#], O. Aberle, F. Cerutti, K. Cornelis, B. Goddard, V. Kain, R. Losito, F. L. Maciariello, M. Meddahi, J. Uythoven, F. M. Velotti, CERN, Geneva, Switzerland,
E. Gianfelice-Wendt, Fermilab[*], Batavia, USA
A. Mereghetti, UMAN, Manchester, UK, and CERN, Geneva, Switzerland

*Abstract*

The challenging High Luminosity LHC (HL-LHC) beam requirements will lead in the future to unprecedented beam parameters along the LHC injector chain. In the SPS accelerator these requests translate into about a factor two higher intensity and brightness than the present design performance. In addition to the challenge of producing and accelerating such beams, these parameters affect the resistance of the existing equipment against beam impact. Most of the protection devices in the SPS ring, its transfer lines and the LHC injection areas will be put under operational constraints which are beyond their design specification. The equipment concerned has been reviewed and their resistance to the HL-LHC beams checked. Theoretical and simulation studies have been performed for the SPS beam scraping system, the protection devices and the dump absorbers of the SPS-to-LHC transfer lines, as well as for the LHC injection protection devices. The first results of these studies are reported, together with the future prospects.

## INTRODUCTION

The LHC Injectors Upgrade project, hereafter LIU [1], is responsible for reliably delivering the required HL-LHC beams. The LHC injector chain includes the LINAC4, PS booster, PS, SPS, as well as the heavy ion chain. The HL-LHC beam specification at 7 TeV (25 ns bunch spacing, $2.2\times10^{11}$ p/bunch within 2.5 μm emittance [2]) translates into challenging beam performances along the injector chain, as illustrated in Table 1 (at the exit of the SPS).

Table 1: 450 GeV SPS Beam Specifications for 25 ns "Nominal" and "Maximum" Beam Parameters

| Parameter | LIU Nominal | LIU Maximum |
|---|---|---|
| Bunch Intensity ($\times 10^{11}$ p) | 1.15 | 2.5 |
| # Bunches | 288 | 288 |
| $\varepsilon^N_{x,y}$ (μm) | 1 | 2.5 |
| $4\sigma_t$ (ns) | 1 | 4 |
| $\sigma_{\Delta p/p}$ ($\times 10^{-3}$) | 0.131 | 0.525 |

$\varepsilon^N_{x,y}$ is the normalised transverse emittance; $\sigma_t$ is the bunch length; $\sigma_{\Delta p/p}$ standard deviation of the momentum spread.

[#]oznur.mete@cern.ch
[*] Operated by Fermi Research Alliance, LLC under Contract No. DE-AC02-07CH11359 with the United States Department of Energy.

Within the LIU project all protection devices and beam dump systems of the SPS accelerator, the SPS-to-LHC transfer lines and the LHC injection regions are being revised in view of the high brightness beam operation.

## NEW SPS BEAM CLEANING CONCEPT

In order to reduce beam losses at the transfer line collimators and the LHC entrance, the LHC beam halo is scraped in the SPS by means of movable blades. The performance of the installed system has been compared to a new system composed of a fixed absorber block onto which the beam is positioned using a closed magnetic bump [3]. The feasibility tests performed during SPS Machine Development sessions [4] proved that scraping with a fixed mask is efficient (Fig. 1). It has also the advantage of allowing scraping at any time in the SPS cycle and being useable as a beam profile monitor (Fig. 2). A review was organised which recommended on the future choice of the SPS scraping system [5]. It was suggested to continue operating with the present system, while performing additional mechanical fatigue tests and evaluating the thermo-mechanical resistance when 25% of the full beam is scraped away. In addition, further new possible material and designs of the blade will be studied. Additional protection devices in the vicinity of the scraping system will be assessed. The new proposed scraping system will be documented in a Technical Design Report, for rapid implementation in case of future need [6].

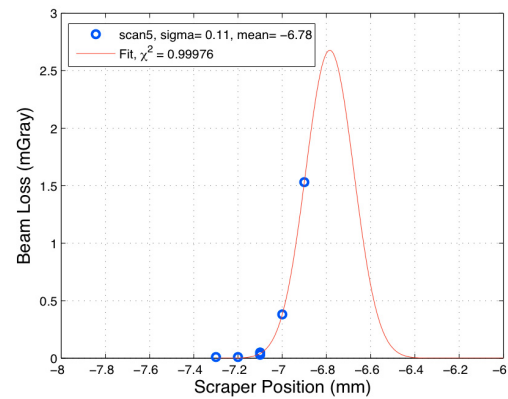

Figure 1: Beam profile acquired with the present movable scraper after the beam cleaning with fixed-scraper in the LSS6 demonstrating that the tails are successfully scraped away with the new method.

The FLUKA and SixTrack codes have been coupled [7] to estimate energy deposition levels in the new scraper. Studies with 4D tracking were performed for full scraping of the maximum LIU beam (Table 1) as worst scenario, moving a 1 m long fixed absorber, made of a carbon fibre composite (CFC, 1.7 g cm$^{-3}$) [8], towards the beam. In addition, the magnetic bump rise time, absorber length and impact angle were varied in order to check dependencies [9]. The maximum energy deposition for all considered cases ranges between 10 and 60 kJ cm$^{-3}$ per bunch train. For comparison, 200 kJ cm$^{-3}$ per bunch train are expected as maximum value of energy deposition in the present scraper, for the same worst case. The computational tools are being further developed to properly simulate the rising of the bump and enable the coupling in 6D tracking [7].

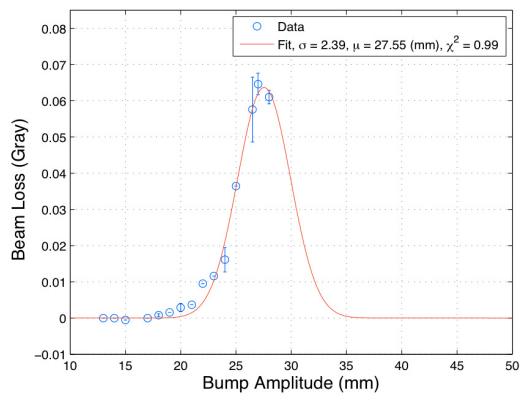

Figure 2: Horizontal beam profile scan using a fixed mask and extraction bumpers in the LSS6 during the SPS machine studies.

## PROTECTION DEVICES

### Protection Devices in the SPS and the SPS-to-LHC Transfer Lines

In the frame of the LIU-SPS activities, the protection devices and beam obstacles were reviewed to address potential issues / observations and to propose solutions. This concerns the SPS internal beam dump (TIDVG), the protective shields for extraction septa (TPSG), the SPS-to-LHC transfer line collimators (TCDI), transfer line beam dumps (TED), and all primary beam line windows and protection devices in the LHC injection regions (TDI).

3D energy deposition maps were calculated in FLUKA for different scenarios of direct impact onto TCDI jaws and TED dumps, to feed further thermo-mechanical analyses with the ANSYS code. Though these studies were focussed on the TCDI and TED, the new material proposals can be extended to all protection devices and obstacles, including the scraper blocks.

### SPS Internal Beam Dump

The temperature of the present TIDVG can exceed the bake-out temperature during continuous beam dumping. Due to the resulting outgassing and the vacuum pressure rise in the upstream injection kickers (MKPs), the dump system is being reviewed, considering as well a possible external dump, the improvement of the dump kicker (MKDV) and the TIDVG vacuum performance [10].

### SPS-to-LHC Transfer Line Collimators

The TCDI locations in the SPS-to-LHC transfer lines were updated, studying the levels of the beam loss showers during HL-LHC beam operation. The study led to a proposal for new collimator locations in the transfer lines [11].

The energy deposition (25 ns bunch spacing, i.e. 288 bunches) on the TCDI jaws (1200x28x60 mm, made of graphite) was studied for nominal and maximum LIU parameters and for different SPS optics [12]. The maximum peak energy deposition is 4.4 kJ $cm^{-3}$ and yields a temperature increase of $1450^oC$. The equipment downstream of TCDIs is protected by masks (TCDIMs). The structural analysis showed that for the TCDIs the main problem is the compression stress with a maximum of 143 MPa, whereas the limit is estimated to be 125 MPa.

### SPS-to-LHC Transfer Line Beam Dumps

The geometry and material configurations of the existing TED were assessed in terms of energy deposition, thermal and structural properties. For the TEDs (4 dumps located in the SPS-to-LHC transfer lines) [13], the worst scenario yields to energy deposition of 3.7 kJ $cm^{-3}$ per bunch train, corresponding to a temperature increase of 1250$^o$ C. Similarly to the TCDI collimator cases, the main structural problem of the present TED concerns the graphite block of its sandwich structure, in which both the compression and tension limits are exceeded.

### Design Studies Relevant for all Protection Devices

New designs in terms of material, geometry and structure capable of withstanding the hottest points and highest thermo-mechanical stresses, are being developed for the protection devices. Different materials (Graphite R4550, Boron Nitride, CFC 1.4, and CFC 1.7) were investigated for their thermal (maximum temperature reached vs. melting point) and structural (maximum compression stress vs. limit in compression) properties (Table 2) [14]. Two methods are used to study thermal-structural problems: weak (or sequential) coupling and strong (or direct) coupling. The latter is also called dynamic analysis and is important for a realistic study. Only a strongly coupled (dynamic) analysis can provide a process history, nevertheless it does not provide a better resolution than the weakly coupled analysis. The current candidate for the TED first range of blocks is CFC at 1.7 g $cm^{-3}$ where the maximum of the energy deposition is expected (Fig. 3). Both analyses (weakly and strongly coupled) were performed for CFC at 1.7 g $cm^{-3}$ resulted in maximum stresses below the limit suggested for the material.

For the next steps, studies will focus on the strongly coupled analysis for the scraper system. The material energy deposition for TCDIs and TEDs will be re-assessed for the chosen material and a design performed.

Table 2: Limits and Structural Properties of Several Materials Under LIU Beam Parameters

|  | Traction | | Compression | |
| --- | --- | --- | --- | --- |
|  | Max.Value [MPa] | Limit [MPa] | Max.Value [MPa] | Limit [MPa] |
| Graphite R4550 | 26 | 38 | 143 | 125 |
| Boron Nitride | 105 | 29 | 700 | 512 |
| CFC 1.4 | 5 | 30 | 45 | 82 |
| CFC 1.7 | 4.8 | 12.8 | 31 | 132 |

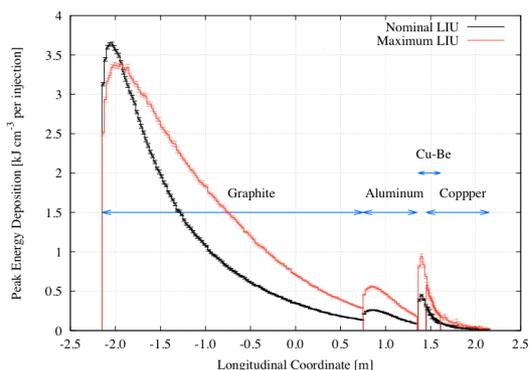

Figure 3: Energy deposition along the TED.

*Protection Devices in LHC Injection Regions*

The LHC is protected against failures of the LHC injection kicker magnets by the injection absorber TDI. In the future the TDI has to be capable to absorb a full injection, of 288 bunches with LIU maximum beam parameters. Beam impact under grazing incidence poses the most stringent requirements to the TDI. The TDI presently installed in the LHC will not be able to absorb this beam. Upgrade studies have started.

The present TDI is a sandwich structure of different materials with a total absorption length of 4.2 m. The upper and lower jaws are located in a single vacuum tank of 5.4 m length. To be able to absorb the LIU beams the total absorption length will need to be increased significantly. Different absorber materials will need to be used based on similar studies as mentioned above for the SPS beam dump.

Significant beam induced heating of the present TDI structure took place during operation. This has led to deformation of the absorber blocks and problems to precisely measure the absorber position. Also for this reason the new TDI will most likely be distributed over several vacuum tanks. An option of 5 tanks with an absorber length of 1.5 m each is presently under study. Great care needs to be taken to reduce the beam impedance of the new TDIs.

It is presently not foreseen to upgrade the auxiliary absorbers TCLIA and TCLIB, placed at different phase advances from the injection kickers, nor the protection masks TCLIM and TCDD.

## CONCLUSIONS AND OUTLOOK

In the frame of the LIU-SPS activities, the protection devices and beam obstacles are being reviewed. A new SPS scraping system was designed, proposed and documented in a technical design report for possible use in future operation. The locations of the transfer line collimators have been optimised. The SPS internal beam dump system is being studied, considering as well a possible external dump, the improvement of the dump kicker (MKDV) and the TIDVG vacuum performance. The existing transfer line beam dumps were also assessed, confirming the need of a new system to withstand the LIU beam. New designs in terms of material, geometry and structure capable of withstanding the hottest points and highest thermo-mechanical stresses, are being developed for all protection devices. The potential issues of the primary beam line vacuum windows will be assessed.

On the LHC side the upgrade studies of the injection absorber are progressing to schedule.

## REFERENCES

[1] LHC Injectors Upgrade Project, Website, https://espace.cern.ch/liu-project/default.aspx
[2] Stretched HL-LHC Baseline parameters, 8 November 2012, private communication with M.Zerlauth and O.Bruning.
[3] O. Mete et al., "Design of a Magnetic Bump Tail Scraping System for the CERN SPS", THPEA040, IPAC2013.
[4] O. Mete et al., "Fixed Scraping Concept" EDMS#1254658 and EDMS#1252467, SPS MD Reports.
[5] LIU-SPS Scraping System Review, https://indico.cern.ch/conferenceTimeTable.py?confId=221617#20130122
[6] Technical Design Report of the LIU-SPS Scraping System (in progress).
[7] A. Mereghetti, et al., "SixTrack-Fluka Active Coupling for the Ipgrade of the SPS Scrapers", MOPWO034, IPAC2013.
[8] F. L. Maciariello, "Thermo-mechanical Analysis of New Materials (for the TED and TCDI) Interacting with High Energy Particle Beam", Master's Thesis (in progress).
[9] A. Mereghetti, "New SPS Scraping System, FLUKA Simulations", LIU-SPS Scraping System Review Meeting.
[10] A. Stadler, "Studies of the SPS internal dump (TIDVG) for current and future proton beams", SPS Upgrade Meeting, 2009.
[11] E. Gianfelice-Wendt, CERN-ATS-Note-2012-095.
[12] A. Mereghetti et al., "Energy Deposition Studies for the Upgrade of the LHC Injection Lines", WEPEA064, IPAC2013.
[13] F. M. Velotti et al., "Performance Improvements of the SPS Internal Dump for the HL-LHC Beam", THPEA041, IPAC2013.
[14] F. L. Maciariello, "New SPS Scraping System, First Dump Block Design", LIU-SPS Scraping System Review Meeting.